\documentclass[showpacs,preprintnumbers,superscriptaddress,
               amsmath,amssymb,prd]{revtex4}
\usepackage[dvips]{graphicx}

\newcommand{\A}{\mathcal{A}}
\newcommand{\D}{\mathcal{D}}
\newcommand{\W}{\mathcal{W}}
\newcommand{\N}{\mathcal{N}}
\newcommand{\Z}{\mathcal{Z}^{\text{DM}}}
\newcommand{\brho}{\bar{\rho}}
\newcommand{\Sym}{S_{\text{YM}}}
\newcommand{\tr}{\text{tr}}

\newcommand{\x}{\text{x}}
\newcommand{\xt}{\boldsymbol{x}}
\newcommand{\yt}{\boldsymbol{y}}
\newcommand{\zt}{\boldsymbol{z}}

\begin{document}


\title{A description of the target wave-function
       encoded in the source terms}
\author{Kenji Fukushima}
\affiliation{RIKEN BNL Research Center, Brookhaven National
             Laboratory, Upton, New York 11973, U.S.A.}
\begin{abstract}
 We argue that the gauge invariant source terms in the formalism of
 the Color Glass Condensate (CGC) should properly describe the target
 wave-function if the microscopic structure is concerned in the regime
 where the parton distribution is not quite dense.  The quantum
 property of color charge density is incorporated in the quantum
 weight function defined with the source terms.  We sketch that the
 logarithmic source terms encompass a meaningful picture of the
 microscopic structure of the target wave-function.
\end{abstract}
\maketitle


\paragraph*{\bf Introduction}

  The formalism of the Color Glass Condensate (CGC) is elaborated
suitably to describe energetic matter in Quantum Chromodynamics
(QCD)~\cite{McLerran:1993ni,McLerran:1993ka,review}.  When the target
material like a nucleus whose dynamics is mainly governed by QCD is
highly boosted along the positive $z$ direction, more and more partons
with small fraction of the longitudinal momentum
$p^+=(p^0+p^z)/\sqrt{2}$ take part in the dynamics.  Such partons are
commonly called the wee partons.  The basic idea of the CGC is as
follows; the wee partons with $p^+$ smaller than a certain separation
scale $\Lambda$ distribute densely at sufficiently high energy that
the wee parton distribution is well coherent and can be described by
the classical field $\A^\mu_a(x)$.  The description is analogous to
classical electromagnetism then once the saturation is reached.  The
wee parton distribution is to be approximated by solving the classical
equations of motion, i.e.\ Yang-Mills-Maxwell equations.  The color
source $\brho^a(x)$ which generates $\A^\mu_a(x)$ is provided by fast
partons, namely, partons with $p^+$ larger than $\Lambda$.

  The name of the CGC is widely understood:  \textit{Color} is the
quantum number relevant to QCD partons.  \textit{Glass} is not
crystalline microscopically but static enough as compared with our
daily time scale.  Likewise, the fast-moving partons are relatively
static seen from the wee partons due to the time dilation.
\textit{Condensate} means that the wee partons are dense and coherent
represented by a classical field like the field-theoretical
description of the Bose-Einstein Condensate (BEC).  We note, however,
that the physics of the CGC is totally different from that of the BEC.
The CGC is not a state of matter, but rather an effective description
of matter.

  The separation scale $\Lambda$ is physically specified by Bjorken-x
which intuitively signifies the longitudinal momentum fraction between the
target's $P^+$ and the parton's $p^+$ in the infinite momentum frame
where $P^+\sim\infty$.  The random distribution of $\brho^a(x)$
depends on $\Lambda$ or x through the weight function
$\W_\Lambda[\brho]$.  The differential equation of $\W_\Lambda[\brho]$
with respect to $\Lambda$ is the evolution equation by which one can
go toward smaller $\Lambda$ and thus smaller x.  The evolution
equation is an essential tool to investigate the small-x physics.  In
the leading-logarithmic approximation the CGC formalism leads us to
the evolution equation known as the
Jalilian-Marian-Iancu-McLerran-Weigert-Leonidov-Kovner (JIMWLK)
equation~\cite{Jalilian-Marian:1997jx,Jalilian-Marian:1997gr,%
Kovner:2000pt,Iancu:2000hn,Iancu:2001ad,Ferreiro:2001qy,%
Weigert:2000gi}.  Although the JIMWLK equation is successful when the
color charge density $\brho^a(x)$ is as large as $\sim 1/g$ in the
saturation region of $Q^2$ and x, it has turned out that an important
process in lower density regimes is missing in the JIMWLK dynamics.
That is the fluctuation of the parton (gluon) number in the target
wave-function. The JIMWLK equation correctly grasps the saturation
mechanism by the merging process of gluons, while it fails in
describing the splitting process that is indispensable in the dilute
regime~\cite{Iancu:2004es,Iancu:2004iy,Mueller:2005ut}.

  It is obvious what lacks in the JIMWLK dynamics.  If the target
material of the charge density $\brho^a(x)$ travels through the gauge
field $\A^\mu_a(x)$ created by the projectile, then the scattering
amplitude earns an eikonal phase $\sim \exp[-i\brho^a(x)\A^-_a(x)]$,
that means there is a sort of uncertainty principle between the target
and the projectile states.  Thus, if the eigenstate of the target
wave-function is chosen represented in terms of $\brho^a(x)$, the
projectile wave-function is uncertain represented in terms of
$i\delta/\delta\brho^a(x)$, which is analogous to
$\hat{x}|x\rangle=x|x\rangle$ and
$\hat{p}|x\rangle=i\hbar\partial/\partial x|x\rangle$ in Quantum
Mechanics.  The JIMWLK equation contains the full order in
$\brho^a(x)$ but only the first and second order in
$i\delta/\delta\brho^a(x)$ under the assumption that $\brho^a(x)$ is
large (and thus $i\delta/\delta\brho^a(x)$ is small).

  The higher order term in $i\delta/\delta\brho^a(x)$ has been added
in a way consistent with the color dipole model (DM) in
Refs.~\cite{Iancu:2004iy,Mueller:2005ut,Blaizot:2005vf} and the
general evolution kernel or the Hamiltonian including full order in
both $\brho^a(x)$ and $i\delta/\delta\brho^a(x)$ has been achieved in
Refs.~\cite{Levin:2005au,Kovner:2005jc,Hatta:2005rn,Balitsky:2005we}.
The state evolving according to the Hamiltonian should be specified to
write down an evolution equation. Thus,
Refs.~\cite{Kovner:2005jc,Kovner:2005uw,Hatta:2005ia} are seminal
works in which they showed that the DM weight function evolves
properly according to the general Hamiltonian in the dilute regime
(Bremsstrahlung Hamiltonian) in the large $N_c$ limit.

  It is not quite clear, however, how we can push the conventional
strategy to derive the JIMWLK equation toward a lower density region.
Apparently, for one example, the derivation of the general Hamiltonian
in Ref.~\cite{Hatta:2005rn} does not rely on any information of the
target nor projectile wave-function, while the JIMWLK equation deals
with the evolution of the target wave-function.  Its clear
manifestation is;  when one derives the JIMWLK equation, one needs to
have the gauge invariant source terms that induce complicated and
non-local vertices.  In contrast, the general Hamiltonian shows itself
without complicated vertices associated with the source action at all.
This seemingly mystical simplicity as compared with the derivation of
the JIMWLK equation is to be understood as
follows~\cite{Fukushima:2005kk,Fukushima:2006cj}.  The source terms
should carry the information of the target wave-function.  The general
Hamiltonian works when it acts either on the target wave-function that
is embodied as the weight function with the source terms as we will
see, or on the projectile wave-function that is supplied as an
operator to be evaluated with the weight function.  Even though the
Hamiltonian is reduced to the JIMWLK kernel in the dense case, it is
not the JIMWLK equation until the weight function is specified.  Since
the parton distribution is highly dense in the JIMWLK problem, the
concrete form of the source terms is not relevant to the final result.
In the dilute regime, however, the explicit form is significant as
demonstrated in the choice of the weight function as in the DM in
Ref.~\cite{Hatta:2005ia}.  This point of view is already implied in a
slightly different formulation of the high-energy evolution in
Refs.~\cite{Kovner:2005jc,Kovner:2005uw}, while we will address here
how we should understand this along with rather conventional JIMWLK
formulations~\cite{Iancu:2000hn,Iancu:2001ad,Ferreiro:2001qy}.

  We will start with the generating functional and briefly review how
the CGC picture arises from the stationary-point approximation.  Then,
we illustrate a simple algebraic trick proposed by the present author
in Refs.~\cite{Fukushima:2005kk,Fukushima:2006cj} to treat the quantum
nature of color charge density.  The benefit of this trick is also
that the physical meaning of the source terms is transparent.
Actually it becomes clear that the source terms should describe the
target wave-function, that is, in the dilute regime the source terms
should comply with the microscopic structure of the target
wave-function.  We will claim that the logarithmic form would be a
likely candidate.
\vspace{2mm}


\paragraph*{\bf Stationary-point approximation}

  In the presence of the color source that distributes with the weight
function $\W_\tau[\brho]$, the QCD generating functional is modeled as
\begin{equation}
 Z[j]=\int\!\D\brho\,\W_\tau[\brho]\int^\tau\!\!\D A \,\delta[A^+]\,
  \exp\biggl\{ i\Sym[A]+iS_W[\brho] -i\!\int\!d^4x\,
  j_\mu^a(x) A^\mu_a(x)\biggr\}
\label{eq:generating}
\end{equation}
in the light-cone gauge.  Here $\Sym[A]$ is the Yang-Mills action and
$S_W[\brho]$ denotes the source action which is given in terms of the
gauge invariant variable,
\begin{equation}
 W[A^-](x^-,\xt) = \mathcal{P}_{x^+}\exp\biggl\{ig
  \int_{-\infty}^{\infty}\!\! dx^+ A^-(x^+,x^-,\xt) \biggr\},
\label{eq:w}
\end{equation}
when the background field comes only from the target moving along the
$x^+$ direction.   It should be noted that $A^-$ in the above
expression is a color matrix either in the fundamental or adjoint
representation in accord with the constituent of the target.  It is
the central issue of our interest how we should fix the concrete form
of $S_W[A^-]$.  In writing Eq.~(\ref{eq:generating}) we introduced the
rapidity variable, $\tau=\ln[P^+/\Lambda]=\ln[1/\x]$, instead of the
separation scale $\Lambda$ or Bjorken-x.  Then, one can compute the
expectation value of an operator $\mathcal{O}[A]$ as
\begin{equation}
 \langle\mathcal{O}[A]\rangle_\tau = \mathcal{O}[i\,\delta/\delta j]
  \,Z[j]\, \bigr|_{j=0}\, /\,Z[0].
\label{eq:expectation}
\end{equation}

  The CGC results from the stationary-point approximation on
Eq.~(\ref{eq:generating}).  When $\brho^a(x)$ is as large as
$\sim 1/g$ at small x, the classical approximation works well, so that
the functional integral over $A_\mu^a$ can be estimated at the
stationary point determined by
\begin{equation}
 \biggl(\frac{\delta\Sym}{\delta A_\mu^a}
  +\frac{\delta S_W}{\delta A_\mu^a}\biggr)_{A=\A} = 0.
\label{eq:eom}
\end{equation}
Then the expectation value (\ref{eq:expectation}) is approximated as
\begin{equation}
 \langle\mathcal{O}[A]\rangle_\tau = \int\!\D\brho\,\W_\tau[\brho]\,
  \mathcal{O}\bigl[\A[\brho]\bigr]
\end{equation}
under the assumption that $\mathcal{O}[A]$ would hardly affect the
stationary condition.  In case of computing the scattering amplitude
between a heavy target of the CGC and a simple projectile like the
color dipole, the above stationary-point approximation is acceptable,
while one needs to solve the two-source problem at the classical level
if one intends to compute observables like the gluon or quark
production associated with the
scattering~\cite{Kovner:1995ja,Dumitru:2001ux,Blaizot:2004wu,Blaizot:2004wv}
even though it is a dense-dilute system.  Here we only care about the
target wave-function and pay little attention to the projectile.

  Let us consider what is required for the property of $S_W[\brho]$.
The leading term of $S_W[\brho]$ should reproduce the eikonal coupling
$-\int\!d^4x\,\brho^a(\vec{x})A^-_a(x)$.  From the equations of motion
(\ref{eq:eom}) the charge density inferred from the source terms
should satisfy the covariant conservation, which is fulfilled whenever
the source terms are written in terms of the matrix elements of
Eq.~(\ref{eq:w}) as one can easily confirm
\begin{equation}
 \begin{split}
 & \bigl(\partial^-\delta^{ab}+gf^{acb}A^-_c\bigr) J^+_b = 0,\\
 & J^+_b \propto \bigl(W_{\infty,x^+}\, ig\, T^b\, W_{x^+,-\infty}
  \bigr)_{ij}.
 \end{split}
\label{eq:conservation}
\end{equation}
Here $T^b$ is an element of the SU($N_c$) algebra.  The covariant
conservation holds for both the fundamental and the adjoint Wilson
lines.  In Eq.~(\ref{eq:conservation}) we denoted the Wilson lines
with the $x^+$ integration ranging from $x^+$ to $\infty$ and from
$-\infty$ to $x^+$ as $W_{\infty,x^+}$ and $W_{x^+,-\infty}$
respectively.

  In the dense regime there is no further restraint onto the choice of
$S_W[\brho]$ and, though it is not rigorously proved yet, there is
likely a sort of universality and the concrete form of $S_W[\brho]$ is
not relevant to the evolution equation, i.e., the JIMWLK equation.

  A further constraint appears in the dilute regime where the color
charge distribution in the target is not quite dense.  For that
purpose it is convenient to transform the generating functional
(\ref{eq:generating}) into the following representation;
\begin{equation}
 Z[j]=\int\!\D\rho\,\N_\tau[\rho]\int^\tau\!\!\D A\,\delta[A^+]\,
  \exp\biggl\{ i\Sym[A]-i\!\int\!d^4x\,\rho^a(x)A^-_a(x)
  -i\!\int\!d^4x\, j_\mu^a(x) A^\mu_a(x) \biggr\},
\label{eq:rewriting}
\end{equation}
where
\begin{equation}
 \mathcal{N}_\tau[\rho] = \int\!\D\brho\,\W_\tau[\brho]\,
  \exp\Bigl\{ iS_W[-i\,\delta/\delta\rho,\brho] \Bigr\}\;\delta[\rho].
\label{eq:density}
\end{equation}
This is just a simple rewriting, and nevertheless, it can bridge the
formalism toward a DM-like picture as we will see.  Later we will
closely discuss the physical meaning of the expression
(\ref{eq:density}).

  It is a notable feature of this representation that one can see the
non-commutative nature of color charge density manifestly.  The
stationary-point of Eq.~(\ref{eq:rewriting}) is determined by the
equations of motion;
\begin{equation}
 \biggl( \frac{\delta\Sym}{\delta A_\mu^a} - \delta^{\mu -}
  \rho^a \biggr)_{A=\A} = 0,
\end{equation}
and the expectation value of $\mathcal{O}[A]$ is then approximated as
\begin{equation}
 \langle\mathcal{O}[A]\rangle_\tau = \int\!\D\rho\,\N_\tau[\rho]\,
  \mathcal{O}\bigl[\A[\rho]\bigr].
\label{eq:expect_n}
\end{equation}
It is important to note that $\rho^a(x)$ is dependent on $x^+$, which
actually takes care of the full account of $\delta S_W/\delta A_\mu^a$
in Eq.~(\ref{eq:eom}).  Therefore, we can no longer assume that the
color source $\rho^a(x)$ is static once the non-commutativity plays a
role.  In the functional integral formalism we have the commutation
relation as follows;
\begin{align}
 &\bigl\langle \rho^a(x^+,\vec{x})\rho^b(x^+\!-\eta,\vec{y})
  -\rho^b(x^+,\vec{y})\rho^a(x^+\!-\eta,\vec{x})
  \bigr\rangle_{\eta\to 0^+} \notag\\
 =& \int\!\D\rho\,\N_\tau[\rho]\,\bigl\{ \rho^a(x^+,\vec{x})
  \rho^b(x^+\!-\eta,\vec{y})-\rho^b(x^+,\vec{y})
  \rho^a(x^+\!-\eta,\vec{x})\bigr\}_{\eta\to 0^+} \notag\\
 =& -(ig)^2\int\!\D\rho\,\N_\tau[\rho]\,\frac{\delta iS_W}{\delta
  W_{ij}}\,\bigl( W_{\infty,x^+}\,[T^a, T^b]\,W_{x^+,-\infty}
  \bigr)_{ij}\,\delta^{(3)}(\vec{x}-\vec{y}) \notag\\
 =& -igf^{abc}\cdot i\,(ig)\int\!\D\rho\,\N_\tau[\rho]\,
  \frac{\delta iS_W}{\delta W_{ij}}\,\bigl(W_{\infty,x^+}\,T^c\,
  W_{x^+,-\infty}\bigr)_{ij}\,\delta^{(3)}(\vec{x}-\vec{y})
  \notag\\
 =& -igf^{abc}\langle\rho^c(x^+)\rangle\,\delta^{(3)}(\vec{x}-\vec{y}).
\label{eq:commutation}
\end{align}
Here we utilized the following relation,
\begin{equation}
 \int\!dx\, f\Bigl(-i\frac{d}{dx}\Bigr)\,\delta(x)\,x
  = \int\!dx\,\Bigl\{f(0)-i\,f'(0)\frac{d}{dx}-\cdots\Bigr\}\,
  \delta(x)\,x = i\,f'(0).
\end{equation}
Strictly speaking, in fact, the right hand side of
Eq.~(\ref{eq:commutation}) is zero after average.  One can insert an
appropriate operator to make the expectation value nonvanishing and
thus Eq.~(\ref{eq:commutation}) should be understood in such a sense.
What we have seen here is essentially equivalent with the argument by
Kovner and Lublinsky in Ref.~\cite{Kovner:2005aq}, though they just
mentioned a similar form encountered in the DM as a motivation.  We
emphasize that the argument so far is applicable for any $S_W[\brho]$
as long as $S_W[\brho]$ is a function of the Wilson line (\ref{eq:w}).
In the previous works~\cite{Fukushima:2005kk,Fukushima:2006cj} the
present author called $\N_\tau[\rho]$ the density of states.  Here it
would make sense to call $\N_\tau[\rho]$ the quantum weight function
since it properly handles the quantization relation.

  Let us shortly comment on a symmetric representation of the
scattering problem of light materials.  In that case, as the target is
a right mover, we shall denote the above $\brho^a(x)$,
$\W_\tau[\brho]$, and $S_W[\brho]$ as $\brho_{\text{R}}^a(x)$,
$\W_\tau^{\text{R}}[\brho_{\text{R}}]$, and
$S_W^{\text{R}}[\brho_{\text{R}}]$ specifically.  In general the
wave-function of the left-moving projectile can be expressed as
\begin{align}
 \mathcal{O}[A] &=\int\!\D\brho_{\text{L}}\,\W_{Y-\tau}^{\text{L}}
  [\brho_{\text{L}}]\,\exp\Bigl\{iS_W^{\text{L}}
  [A^+,\brho_{\text{L}}]\Bigr\}
\label{eq:op_a} \\
 &=\int\!\D\rho_{\text{L}}\D\brho_{\text{L}}\,\W_{Y-\tau}^{\text{L}}
  [\brho_{\text{L}}]\,\exp\biggl\{iS_W^{\text{L}}
  [-i\delta/\delta\rho_{\text{L}},\brho_{\text{L}}] -i\int\!d^4x\,
  \rho^a_{\text{L}}(x)A^+_a(x)\biggl\}\,\delta[\rho_{\text{L}}]
\label{eq:op_b}
\end{align}
with $Y$ being the total relative rapidity of the target and
projectile.  Then, if the scattering amplitude is concerned and $A^+$
in Eq.~(\ref{eq:op_a}) is simply approximated as $\A^+[\rho]$ like in
Eq.~(\ref{eq:expect_n}), the scattering amplitude is evaluated as
\begin{equation}
 S_Y= \int\!\D\rho\,\D\brho_{\text{R}}\,\D\brho_{\text{L}}\,
  \W_\tau^{\text{R}}[\brho_{\text{R}}]\,\W_{Y-\tau}^{\text{L}}
  [\brho_{\text{L}}]\,\exp\Bigl\{iS_W^{\text{R}}
  [-i\delta/\delta\rho,\brho_{\text{R}}]+iS_W^{\text{L}}
  [\A[\rho],\brho_{\text{L}}]\Bigr\}\,\delta[\rho].
\end{equation}
This expression makes sense in the dilute-dilute or dense-dilute
(where the dense-dilute duality~\cite{Kovner:2005en} realizes)
scatterings, but not in the dense-dense case in which the classical
solution $\A_a^+[\rho]$ must be evaluated in the presence of two dense
sources of the target and projectile.  The color glass representation
of the onium-onium scattering like as formulated in
Ref.~\cite{Iancu:2003uh} is one manifestation of the above.  If one is
interested in an expectation value of some other operators, such as
in the gluon production~\cite{Blaizot:2004wu} and in the quark
production~\cite{Blaizot:2004wv} and so on, or if one is interested in
the dense-dense case, one has to use Eq.~(\ref{eq:op_b}) and solve the
two-source problem with $\rho_{\text{R}}(x)$ and $\rho_{\text{L}}(x)$
as analyzed in
Refs.~\cite{Kovner:1995ja,Dumitru:2001ux,Blaizot:2004wu,Blaizot:2004wv,Krasnitz:1999wc,Krasnitz:2001qu,Krasnitz:2002mn,Lappi:2003bi}.
Now let us return to the main stream of this work and discuss the
choice of the source terms.
\vspace{2mm}


\paragraph*{\bf Choice of the source terms}

  We know that in the large $N_c$ limit the DM weight function
captures the correct description of the target wave-function in the
dilute regime.  The dressed dipole weight function
reads~\cite{Mueller:2005ut,Iancu:2003uh,Hatta:2005ia}
\begin{equation}
 \Z_\tau[\rho] = \sum_{N=1}^\infty \int d\Gamma_N\,
  P_N^{\text{DM}}(\{\zt_i\};\tau) \prod_{i=1}^N
  D^\dagger(\zt_{i-1},\zt_i)\,\delta[\rho],
\label{eq:DM}
\end{equation}
where $N$ is the number of dipoles and $d\Gamma_N$ means the
integration over all the phase space, namely,
$d\Gamma_N=d\zt_1\cdots d\zt_N$ and $D^\dagger(\xt,\yt)$ is the
creation operator of a dressed dipole located at $\xt$ and $\yt$ in
transverse space, that is,
\begin{equation}
 D^\dagger(\xt,\yt)=\frac{1}{N_c}
  \tr\bigl[W(\xt)W^\dagger(\yt)\bigr]
\label{eq:DMop}
\end{equation}
in the fundamental representation.  In the DM the longitudinal
structure in the $x^-$ direction is neglected and the Wilson lines do
not depend on the longitudinal coordinate.  Now one may well notice
that Eq.~(\ref{eq:DM}) is a special case of the quantum weight
function (\ref{eq:density}).  In other words, the source terms
$S_W[A^-]$ must be chosen such that the quantum weight function can
reproduce Eq.~(\ref{eq:DM}) somehow in the large $N_c$ limit;
\begin{equation}
 \N_\tau[\rho]\; \longrightarrow\; \Z_\tau[\rho]
  \hspace{5mm} \text{as\ \ $N_c\to\infty$}.
\label{eq:requirement}
\end{equation}
It is important to mention that we always mean the \textit{dressed}
dipoles written in terms of the Wilson line.  Actually, in the weak
field limit, any source terms which satisfy
$S_W[A]=-\int\!d^3x\,\brho^a(\vec{x})A^-_a(x)+\cdots$ would lead to
the results of the (undressed) DM if only an ensemble of the
dipole-type color source,
\begin{equation}
 \brho^a(\vec{x})=Q^a\delta(x^-)\{\delta^{(2)}(\xt-\xt_0)-
  \delta^{(2)}(\xt-\yt_0)\},
\label{eq:random}
\end{equation}
where $Q^a$ is a Gaussian random variable~\cite{Iancu:2003uh}, is
allowed by $\W_\tau[\brho]$.

  The conventional choice adopted in
literatures~\cite{Jalilian-Marian:1997jx,Jalilian-Marian:1997gr,%
Kovner:2000pt,Iancu:2000hn,Iancu:2001ad,Ferreiro:2001qy,%
Weigert:2000gi}
\begin{equation}
 S_W[\brho] = \frac{i}{gN_c}\int\!d^3x\, \tr\bigl\{
  \brho(\vec{x})\,W[A^-](\vec{x})\bigr\},
\label{eq:conventional}
\end{equation}
does not meet the requirement of Eq.~(\ref{eq:requirement}); whatever
$\W_\tau[\brho]$ cannot lead to the form of the weight function
(\ref{eq:DM}).  Therefore, the conventional form of the source terms
is no longer a right choice microscopically and is ruled out.  It is
instructive to look further into why the conventional choice is not
good.  If the above conventional source terms are substituted into the
weight function (\ref{eq:density}) and the exponential is expanded, an
arbitrary number of $W$ is allowed at each point.  The higher order
$W$ terms cannot be controlled by the distribution $\W_\tau[\brho]$
and they are not compatible to the specific form of (\ref{eq:DM}).  To
put it in a more formal way, this incompatibility seems to stem from
that the color charge density belonging to the \textit{algebra} and
the Wilson line belonging to the \textit{group} are put on the equal
footing, which is quite unnatural.

  One possibility would be the eikonal coupling with the Wess-Zumino
action~\cite{Kovner:2005nq,Kovner:2005aq,Hatta:2005wp} that takes care
of the non-commutative nature of color charge density just like our
$\N_\tau[\rho]$.  The Wess-Zumino action $S_{\text{WZ}}[\rho]$ is a
function of $\rho^a(x)$.  In principle $S_{\text{WZ}}[\rho]$ is given
in terms of $W[-i\delta/\delta\rho]$, but in fact,
$W$ is implicit in the expression of $S_{\text{WZ}}[\rho]$ given in
Refs.~\cite{Kovner:2005nq,Hatta:2005wp} and it is non-trivial how to
retrieve $W$ out of $S_{\text{WZ}}[\rho]$.  We have no idea how to
derive the microscopic information of the target wave-function, not
going back to the representation as a product of $W$ but directly
using $S_{\text{WZ}}[\rho]$ on its own.

  Instead, we shall here focus on another specific possibility of the
source terms.  We will check whether the following logarithmic
form~\cite{Fukushima:2005kk,Fukushima:2006cj,Jalilian-Marian:2000ad},
\begin{equation}
 S_W[A^-,\brho] = i\,\frac{2}{g}\int\!d^3x\,\tr\bigl\{\brho(\vec{x})
  \ln W[A^-](\vec{x})\bigr\},
\label{eq:log_source}
\end{equation}
in the fundamental representation, or
\begin{equation}
 S_W[A^-,\brho] = i\,\frac{1}{gN_c}\int\!d^3x\,\tr\bigl\{\brho(\vec{x})
  \ln \widetilde{W}[A^-](\vec{x})\bigr\},
\label{eq:log_source_ad}
\end{equation}
in the adjoint representation ($\widetilde{W}$ is defined with the
adjoint generators, and so is $\brho(\vec{x})$ implicitly), is
qualified or not as a microscopically meaningful choice.  In this form
we note that $\ln W$ belongs to the algebra unlike $W$, so that the
coupling to $\brho^a(x)$ looks reasonable.

  We shall develop one more argument besides
Refs.~\cite{Fukushima:2005kk,Fukushima:2006cj,Jalilian-Marian:2000ad}
that the logarithmic form (\ref{eq:log_source}) is motivated.  We know
that the solution of the covariant conservation
(\ref{eq:conservation}) takes a form of
\begin{equation}
 J^+_b(x^+,\vec{x}) = \widetilde{W}_{x^+,-\infty}^{ba}
  \brho^a_{-\infty}
\end{equation}
with the initial condition being $\brho^a_{-\infty}$ at infinite past.
Using the relations $\widetilde{W}^{ba}=2\tr[t^b W t^a W^\dagger]$ and
$W^\dagger_{x^+,-\infty}=W^\dagger_{\infty,-\infty}W_{\infty,x^+}$, we
can equivalently rewrite the above as
\begin{equation}
 J^+_b = -i\,\frac{2}{g}\tr\biggl[\brho_{-\infty}W^\dagger
  \frac{\delta W}{\delta A^-_b}\biggr]
 =i\,\frac{2}{g}\sum_{n=0}^\infty\tr\biggl[\brho_{-\infty}(1-W)^n
  \frac{\delta(1-W)}{\delta A^-_b}\biggr],
\label{eq:current}
\end{equation}
where we expanded $W^\dagger=\{1-(1-W)\}^{-1}$ in terms of $1-W$.
Because $1-W$ is at least linear in $A^-$ if expanded, intuitively,
$(1-W)^n$ in the above represents the (at least) $n$ gluon insertion.
Now we change the variable $\brho_{-\infty}$ so as to allow for all
the orderings of the source among the gluon insertion, namely,
\begin{equation}
 \brho_{-\infty}(1-W)^n \;\longrightarrow\;
  \frac{1}{n}\Bigl\{\brho\,(1-W)^n + (1-W)\,\brho\,(1-W)^{n-1}
  +(1-W)^2\,\brho\,(1-W)^{n-2} +\cdots\Bigr\}.
\label{eq:change}
\end{equation}
Then, Eq.~(\ref{eq:current}) simplifies as
\begin{equation}
 J^+_b=-i\,\frac{2}{g}\tr\biggl[\brho\,\frac{\delta \ln W}
  {\delta A^-_b}\biggr],
\end{equation}
which exactly coincides with the current deduced from the logarithmic
source action (\ref{eq:log_source}).  From this argument, it is also
clear that our $\brho(\vec{x})$ does not correspond to the initial
condition at infinite past.  We did not derive
Eq.~(\ref{eq:log_source}) from Eq.~(\ref{eq:conservation}), for
Eq.(\ref{eq:change}) is a highly non-trivial variable change, but we
can insist that Eq.~(\ref{eq:log_source}) rather than the conventional
one is well motivated.

  Returning to the main discussion, let us draw an intuitive picture
of the microscopic description of the target wave-function which
consists of partons propagating along the $x^+$ direction.  The
classical source $\brho^a(x)$ indicates the parton distribution.  It
is quite important to notice that the distribution of the color
orientation of $\brho^a(x)$ is \textit{not} random at each spatial
point.  Of course, in the dense regime, one can regard the
coarse-grained color distribution as random at each point, but one
cannot in the dilute regime.  Otherwise there would be no correlation
between different two points of a quark and an antiquark and the
dipole-type operator like Eq.~(\ref{eq:DMop}) is not permitted but
only $\tr W(\xt)\,\tr W^\dagger(\yt)$ would arise.  The random
distribution is expected to be valid at scale longer than the
saturation scale $\sim Q_s^{-1}$ and the dipole size would be
smaller.  The above mentioned is explicitly observed in
Eq.~(\ref{eq:random}) where the random variable is common for the
quark at $\xt_0$ and the antiquark at $\yt_0$.

  In order to see the meaning of the logarithmic form, for the
simplest example, let us assume that a quark and an antiquark sit at
$\xt_0$ and $\yt_0$ respectively in the same way as in
Eq.~(\ref{eq:random}).  (We will consider the adjoint case later.)  As
we explained above, the spatial resolution of randomness is coarser
than the separation scale of the dipole.  Then, the source part
becomes
\begin{equation}
 \exp\biggl[-\frac{2}{g}\tr\bigl\{\brho\bigl(\ln W(\xt_0)
  +\ln W^\dagger (\yt_0)\bigr)
  \bigr\}\biggr],
\label{eq:two_source}
\end{equation}
where we used the fundamental one (\ref{eq:log_source}) and
$-\ln W=\ln W^\dagger$.  In general, of course, there could be a color
rotation between $\brho^a(\xt_0)$ and $\brho^a(\yt_0)$.  If
$\brho^a(\yt_0)t^a = S\brho^a(\xt_0)t^a S^\dagger$, however, such a
color rotation by $S$ can be absorbed by the redefinition
$S^\dagger W^\dagger(\yt_0) S\to W^\dagger(\yt_0)$.  The point here is
that the random average over $\brho^a$ is taken simultaneously for the
quark and antiquark and Eq.~(\ref{eq:two_source}) is then a natural
extension of Eq.~(\ref{eq:random}) to the non-linear situation.

\begin{figure}
\begin{minipage}{7.cm}
 \includegraphics[width=4cm]{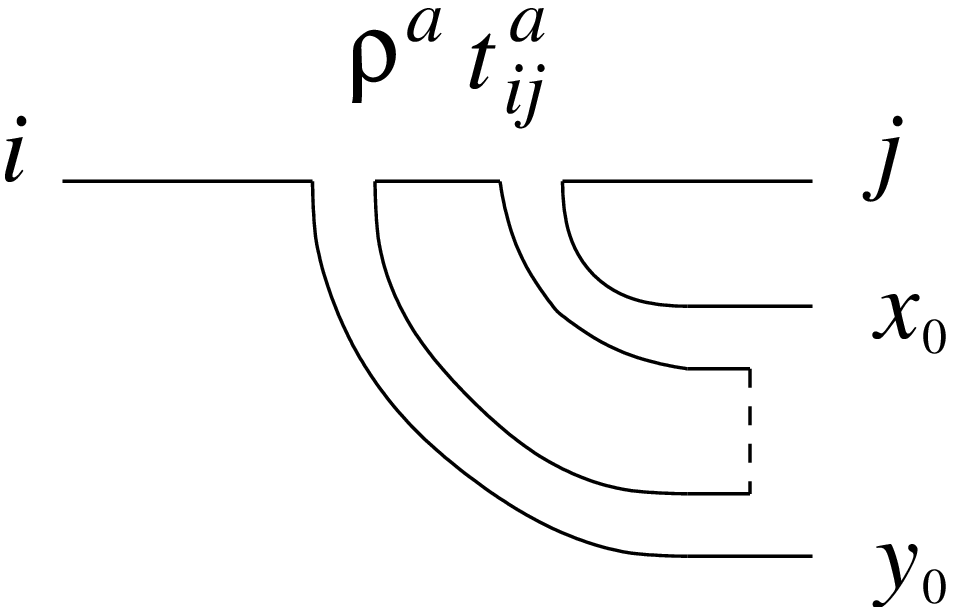}
 \caption{Intuitive picture of the leading contribution in the $N_c$
 counting.}
 \label{fig:quark}
\end{minipage}
\hspace{1cm}
\begin{minipage}{7cm}
 \includegraphics[width=4cm]{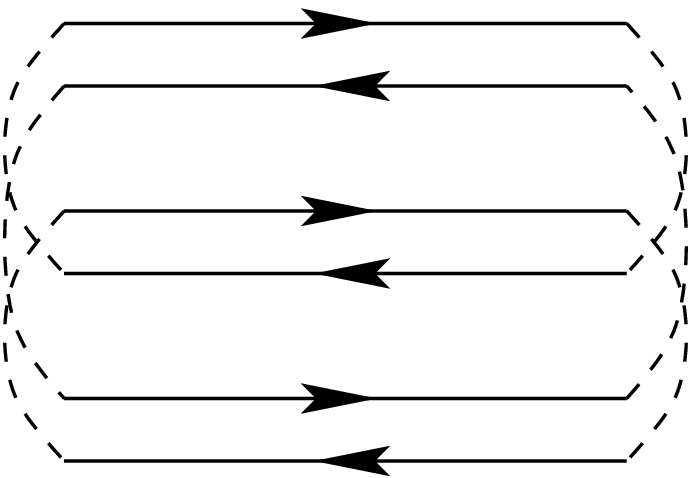}
 \caption{An example of the generalization to the case with multiple
 adjoint particles.}
 \label{fig:gluon}
\end{minipage}
\end{figure}

  Here one can make use of the Campbell-Hausdorff formula,
\begin{equation}
 \ln W(\xt_0) + \ln W^\dagger(\yt_0) = \ln\bigl[ W(\xt_0)
  W^\dagger(\yt_0) \bigr] -\frac{1}{2}\bigl[ \ln W(\xt_0),\;
  \ln W^\dagger(\yt_0) \bigr] +\cdots.
\label{eq:campbell}
\end{equation}
In the large $N_c$ limit only the first term is dominant;  the gauge
coupling scales as $g\sim 1/\sqrt{N_c}$ which is compensated by quark
loops made up in the double line notation of gluons.  The higher order
commutators lead to the higher order terms in $g$ and yet the color
structure is reduced by the commutators, and thus they are suppressed.
At the price of mathematical rigor we sketch the intuitive picture of
this in Fig.~\ref{fig:quark}.  Recalling that the source terms
originate from the QCD vertex $g\bar{\psi}_i A^-_a t^a_{ij}\psi_j$ in
case of the fundamental representation, Eq.~(\ref{eq:two_source}) can
be expressed as the one-gluon vertex between a propagating quark and
gluons as in Fig.~\ref{fig:quark}.  Here we adopted the double line
notation for gluons at $\xt_0$ and $\yt_0$.  Roughly the first term of
Eq.~(\ref{eq:campbell}) corresponds to the contribution with the
dotted line in Fig.~\ref{fig:quark} that makes a loop and thus is the
leading order in the $N_c$ counting.

  We drop the sub-leading commutators in Eq.~(\ref{eq:campbell}).
Therefore, in the presence of a quark and an antiquark and in the
large $N_c$ limit, the quantum weight function reads
\begin{equation}
 \N_\tau[\rho]\sim \int\!d\brho\, \W_\tau[\brho]\,
  \exp\biggl[-\frac{2}{g}\tr\bigl\{\brho\ln\bigl[W(\xt_0)
  W^\dagger(\yt_0)\bigr]\bigr\}\biggr]\,\delta[\rho].
\label{eq:leading}
\end{equation}
Now the calculation goes in the exactly converse manner as in
Ref.~\cite{Fukushima:2005kk}.  From the property,
$\W_\tau[S\brho S^\dagger]=\W_\tau[\brho]$, one can diagonalize
$W(\xt_0)W^\dagger(\yt_0)$ to have a polynomial in terms of its
eigenvalues.  Using $\W_\tau[S\brho S^\dagger]=\W_\tau[\brho]$ again,
one can conclude that the polynomial is to be expressed in terms of
the trace of $W(\xt_0)W^\dagger(\yt_0)$ (see also the discussions in
Ref.~\cite{Fukushima:2005kk}).  Hence, when $\W_\tau[\brho]$ allows
for up to the first order of the trace, we reach
\begin{equation}
 \N_\tau[\rho]\sim D^\dagger(\xt_0,\yt_0)\,\delta[\rho].
\end{equation}
What we showed here is that there \textit{is} a certain
$\W_\tau[\brho]$ that makes the weight function (\ref{eq:density})
with the logarithmic source terms (\ref{eq:log_source}) being the DM
weight function (\ref{eq:DM}), but there is \textit{not} for the case
with the conventional source terms (\ref{eq:conventional}).

  The generalization to the case with more quarks and antiquarks is
straightforward under the assumption that the randomness of color
orientation is clustered at the scale of dipoles.  Also, the
generalization to the adjoint source (\ref{eq:log_source_ad}), i.e.\
the case with the target composed of gluons, is easy too.  The above
argument works as it is with the Wilson line replaced by the adjoint
one $\widetilde{W}$.  The formula,
$t^a_{ij}t^b_{kl}=\frac{1}{2}\delta_{il}\delta_{jk}-\frac{1}{2N_c}
\delta_{ij}\delta_{kl}$, follows that the leading order at large $N_c$
is given by all the dipole-type combinations of gluon double lines, as
exemplified in Fig.~\ref{fig:gluon}.  Thus, as an ensemble, these
combinations eventually lead to the DM picture with a certain weight
$P^{\text{DM}}_N$.
\vspace{2mm}


\paragraph*{\bf Discussions}

  We demonstrated that the logarithmic form, at least in principle,
encompasses the desirable microscopic structure of the target
wave-function.  It had been confirmed that the logarithmic source
terms correctly lead to the JIMWLK equation in the dense regime, and
in addition to that, it has been checked here that they are qualified
also in the dilute regime.  In fact, however, the logarithmic form of
the source terms alone does not provide sufficient microscopic
information.  That means, in order to determine the dynamics in the
dilute regime, the microscopic information of $\W_\tau[\brho]$ is
required;  especially the randomness is only at the unit of the dipole
size scale, or in other words, the color orientation has a strong
correlation at short distance of the dipole size.  The coarse-grained
picture of the color average is valid only in the dense regime.

  Let us assume that we had a complete evolution equation applicable
for the dense, dilute, and intermediate regimes.  It would be an
intriguing question what would transpire at intermediate density.
When the source is large enough in the dense case, the color
distribution is regarded conventionally as random.  That is, in
two-dimensional transverse space, it is completely
\textit{disordered}.  When the density gets lower and lower, the finer
structure of randomness becomes significant, and the two-dimensional
system is no longer disordered and a sort of \textit{localization} is
important.  We have no idea so far what dynamics would separate these
two distinct regimes.  Is it a continuous crossover from one to the
other, or does it result in a discontinuous transition between two
regimes?

  It is known that neither of the JIMWLK and the DM equations would be
a good starting point to tackle this problem in the intermediate
density region.  The JIMWLK dynamics is valid after coarse graining
and thus it contains less information.  The DM might be good, but it
is unlikely that the DM is capable of taking care of saturation, for
which the interaction among partons is important.  We would expect
that our logarithmic expression may give a hint for future
developments.
\vspace{2mm}


\paragraph*{\bf Acknowledgments}

  The author thanks Y.~Hatta for critical discussions and L.~McLerran
for drawing the author's attention to the lower density case.  This
work was supported by the RIKEN BNL Research Center and the U.S.\
Department of Energy (D.O.E.) under cooperative research agreement
\#DE-AC02-98CH10886.


\end{document}